\documentclass[doublecol]{epl2}
\usepackage{graphicx}
\usepackage{amssymb}
\usepackage{amsmath}
\newcommand{\be}{\begin{equation}}
\newcommand{\ee}{\end{equation}}
\newcommand{\ben}{\begin{eqnarray}}
\newcommand{\een}{\end{eqnarray}}
\newcommand{\bes}{\begin{subequations}}
\newcommand{\ees}{\end{subequations}}

\title{Crossover of thermal to shot noise in chaotic cavities}

\author{A. L. R. Barbosa\inst{1} \and  J. G. G. S. Ramos\inst{2} \and D. Bazeia\inst{2}}
\shortauthor{D. Bazeia\etal}

\institute{
\inst{1} EAD-F\'isica, Universidade Federal Rural de Pernambuco, 52171-900 Recife, PE, Brazil\\
\inst{2} Departamento de F\'\i sica, Universidade Federal da Para\'\i ba, 58051-900 Jo\~ao Pessoa, PB, Brazil}


\abstract{We study the crossover between thermal and shot-noise power in a chaotic quantum dot in the presence of non-ideal contacts at finite temperature. The result explicitly demonstrates that the temperature affect the suppression-amplification effect present in the main quantum noise. In particular, the weak localization contribution to the noise has an anomalous thermal behavior when one let the barriers vary, indicating the presence of a critical point related to specific value of the tunneling barriers.
We also show how to get to the opaque limit of the quantum dot at finite temperature.}

\begin{document}
\maketitle


Mesoscopic systems are of great importance to investigate quantum effects at extreme conditions. The presence of external fields, the discreteness of the electric charge and other controllable fundamental physical quantities can directly affect the quantum interference between the electronic modes inside a mesoscopic system \cite{blanter1,martin1}. New effects may spring from the new degrees of freedom, which can be used toward applications such as in the construction of a quantum computer. Since the thermal effects may affect the quantum transport through the system \cite{datta,Nazarovbook,beenakker}, in this work we shall study a mesoscopic apparatus at finite temperature.

The system to be investigated is shown in Fig.~1. It contains two electronic reservoirs at the same temperature $T$, but with different chemical potentials, $\mu_1$ and $\mu_2$, one at the left and the other at the right of the dot. The two reservoirs are connected by ideal leads, and the difference between the two chemical potentials may trigger an electronic current through the mesoscopic system. In Ref.~\cite{pekola}, one investigates the system at finite frequency with ideal contacts. There, the authors study in details the main semiclassical term, without including corrections due to quantum interference. In this Letter, we shall be mainly concerned with properties of the first quantum correction in the presence of barriers, but at zero frequency. Such study is of current interest, since it will show the appearance of competing properties between tunneling and temperature on the term which collects quantum interference. The key result of this study closes a gap in the literature and allows pointing effects not seen before. Moreover, the finite temperature may generate thermal fluctuations in the occupation number at the Fermi level and contribute to the electronic current in the mesoscopic system. This is the thermal noise \cite{Nyquist,Johnson}, and it is also of direct interest, being related to the conductance through the dissipation-fluctuation theorem.

\begin{figure}[h!]
\begin{center}
\includegraphics[width=8.2cm,height=3.0cm]{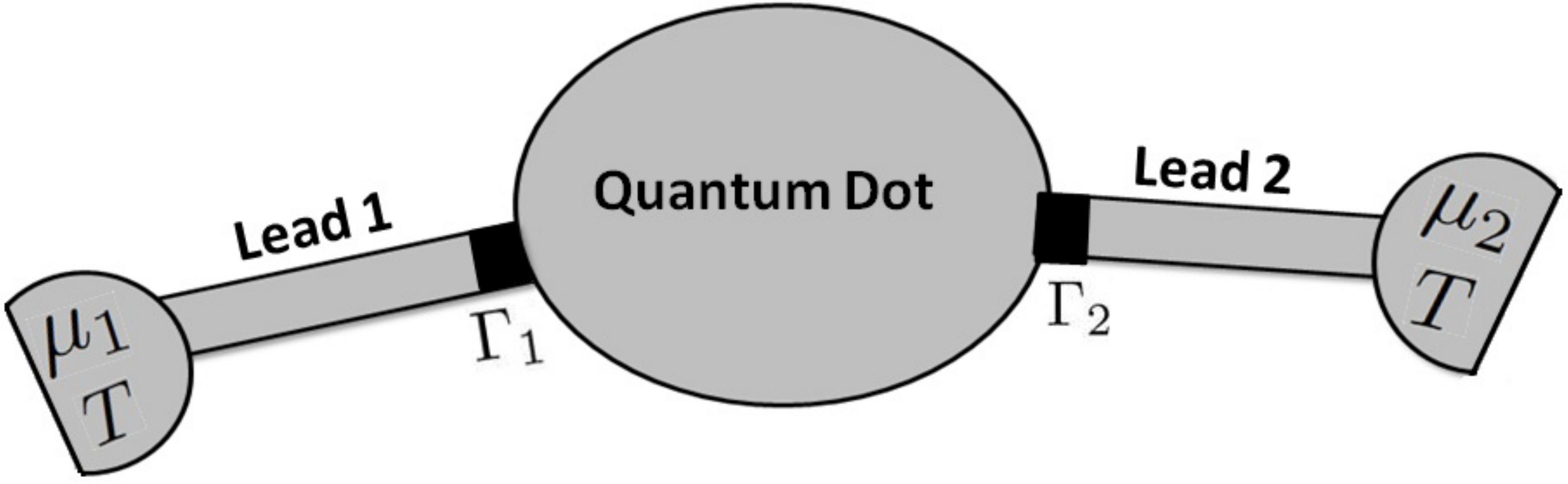}
\end{center}
\caption{Schematic view of the quantum dot.}\label{sistema}
\end{figure}

Another source of noise is associated with the discreteness of the electric charge. If the current through the mesoscopic system is of low intensity, each charge carrier is of great importance, so the number of modes related to the charge carriers in a low intensity flux strongly constrains the transmission process and induces a charge transmission statistics in the mesoscopic system. The electronic counting statistics was first developed in Ref.~\cite{Levitov}, in analogy with the photon counting statistics in quantum optics \cite{Mendel}. The method consists in obtaining the probability distribution function $P_n$ to observe the number $n$ of electrons transferred through the mesoscopic system in a certain time interval. The function $P_n$ can be given in terms of a generating function, associated to the measuring process.

The procedure used to extract the full counting statistics is due to Levitov and Lesovik \cite{Levitov}. It works for several open channels and can be inserted in the scattering formalism of Landauer and B\"uttiker in the linear regime. This formalism can be connected to the random-matrix theory (RMT), in a way appropriate to describe universal statistical properties of observables of transport in open chaotic cavities. The universal description is valid for $\tau_{dw}\gg {\rm max}(\tau_{E},\tau_{erg})$, with $\tau_{dw}$, $\tau_{E}$, and $\tau_{erg}$ being the dwell, Ehrenfest, and ergodic time, respectively. In the case of ideal contacts, it can be described in terms of random scattering matrices in the standard Wigner-Dyson ensembles \cite{mello,beenakker}. These ensembles are classified according to some  fundamental symmetries (time reversal and rotation of fermionic half integer spin) which can or cannot be broken by the chaotic dynamics. They are usually labeled by Dyson's integer index $\beta$ \cite{dyson}: the circular orthogonal ensemble (COE) is applicable to systems engendering both time reversal and rotational symmetries $(\beta=1)$; the circular unitary ensemble (CUE) is valid for systems in the absence of both time reversal and rotational symmetries $(\beta=2)$; and the circular simplectic ensemble (CSE), used for systems in the presence of time reversal, but with broken rotational symmetry $(\beta=4)$.

The basic object of the theory is the scattering matrix. It can always be written in the $2\times 2$ form
\begin{equation}
 S = \left( \begin{array}{cc}
r & t \\
t'& r' \end{array} \right), \label{Smatrix}
\end{equation}
where $r$, $r'$ and $t$, $t'$ are reflection and transmission matrices, respectively. Using the Landauer scattering approach to
quantum transport \cite{datta} we can relate certain moments of the transmission matrix to transport observables. For
instance, the dimensionless conductance (first moment) has the form $\mathbf{g}_1= \mathbf{Tr}(tt^{\dagger})$.
Thus, introducing $\mathbf{g}_{2}= \mathbf{Tr}[(tt^{\dagger})^2],$
we can calculate the shot-noise combining $\mathbf{g}_1$  and $\mathbf{g}_2$ to get \cite{Buttiker1,Buttiker2}
\begin{equation}
\mathbf{p}=\mathbf{g}_1-\mathbf{g}_{2}=
\mathbf{Tr}\left[tt^{\dagger} (1-tt^{\dagger})\right].
\label{ruido}
\end{equation}

If we follow \cite{Blanter}, we see from the main quantum correction to the noise that its value also characterizes interference among the scattering amplitudes. This fact can be better seen as follows: we first realize that the conductance can always be written in terms of the transmission eigenvalues, that is, the conductance can always be written as the sum of the transmission probabilities for each channel separately. And the quantum noise cannot in general be written uniquely in terms of those eigenvalues.

We note that for temperature $T$ much higher than the bias tension $V$, such that $k_{B}T \gg eV$, the current fluctuations are dominated by the thermal noise, also known as the Johnson-Nyquist noise. Here we are using $V=\Delta \mu$, with the bias tension given in terms of the difference between the two chemical potentials $\mu_1$ and $\mu_2$. When the Coulomb interactions can be discarded, the noise is given by ${\cal S}=4k_{B}TG_{0} {\bf Tr}(tt^{\dagger})$, where $G_{0}=e^2/h$ is the conductance quantum. This result was already obtained in \cite{Khlus,levitov1,Buttiker1,Buttiker2,martin}, with the use of the dissipation-fluctuation theorem.

On the other hand, for temperature much lower than the bias tension, $eV \gg k_{B}T$, the current fluctuations are dominated by the shot-noise power. In general, however, at finite temperature both the thermal noise and the shot-noise power contribute to the fluctuations in the current flux. The crossover between the two kind of noise can be obtained from the general expression \cite{Buttiker1,Buttiker2,martin,osipov} in the limit of vanishing frequency
\be
\frac{{\cal S}(k_B T,eV)}{4 k_B T\; G_0 }\!=\!{\bf Tr}(tt^\dag)^2\! +\! F(\theta)\! \left[{\bf Tr}(tt^\dag)-{\bf Tr}(tt^\dag)^2\right], \label{noiseprincipal}
\ee
where $F (\theta)=\theta\coth\left(\theta\right)$ and $ \theta \equiv {eV}/{2k_B T}$.

We will focus mainly on obtaining explicit results for the average of the noise given by the above expression, in the universal regime and in the presence of barriers.
The main quantum correction to the noise plays a key role when the electronic transmission is chaotic; see, e.g., the cases of tunneling junctions or Schottky-like diodes \cite{TH}. Due to its role in the process of charge transmission, we shall also include tunneling barriers in the present procedure, which will also lead us to study opacity of the mesoscopic system.

In the case of chaotic cavity in the universal regime, the main issue is to calculate the average noise. We propose as an efficient way to make the calculation, the diagrammatic procedure developed by Brouwer and Beenakker \cite{bb96}. The procedure consists in calculating average in the unitary ensemble with the Haar measure, together with the Poisson kernel, implementing an expansion in the number of open channels in the leads, known as the semiclassical regime. The two main terms in the expansion are generated by ladder diagrams, leading to the main term, and by maximally crossed diagrams, giving the quantum correction. The main term is known as the diffuson \cite{martin1}, since it can be obtained by means of a diffusion equation. The other maximally crossed terms represent the main quantum interference terms. They are known as cooperons \cite{martin1}, and appear due to the spatial and temporal coherence among the propagating electronic modes.

We introduce an analytical procedure for the case of ideal contacts, for which the Poisson kernel equals the identity and simplifies the calculation, giving rise to non-perturbative result such as the one obtained in Ref.~\cite{bb96}. We focus attention on the unitary group, to calculate the average of the trace of product of unitary matrices, which are the basic elements to lead us to the average of the Eq.~\eqref{noiseprincipal}. Here we get
\begin{eqnarray}
\langle \mathbf{Tr}(tt^\dag)\rangle =\langle \mathbf{Tr}(C_{1}SC_{2}S^\dagger)\rangle=
\frac{N_1N_2}{N-1+\frac{2}{\beta}} \label{condutanciabeta},
\end{eqnarray}
where $\beta\in\{1,2,4\}$ is Dyson's symmetry index. Also, $N_1$ and
$N_2$ stand for the number of open scattering channels in leads 1 and 2,
respectively. Thus, $N=N_1+N_2$ gives the total number of open scattering channels. The  $S$-matrix in this formula can be represented as in
(\ref{Smatrix}) and can be used to describe the chaotic scattering inside the cavity. Of course, it is distributed with $\beta=1,2,$ or $4$, according to the relevant ensemble required by the symmetry of the system under consideration. Moreover, $C_1$ and $C_2$ represent projection matrices defined by
\begin{equation}
 C_1 = \left( \begin{array}{cc}
1_{N_1} & 0 \\
0 & 0 \end{array} \right),\;\;\; C_2 = \left( \begin{array}{cc}
0 & 0 \\
0 & 1_{N_2} \end{array} \right),\nonumber
\end{equation}
in which $1_{N_j}$ is the $N_j\times N_j$ unity matrix. The projection matrices are introduced in Eq.~\eqref{condutanciabeta} to make the integration on unitary matrices. We recall that the $S$ matrix is unitary due to current conservation. They obey $C_1 C_2=0$ and $C_1+C_2=1_N$. The diagrammatic method generates ciclic permutations of the matrix elements, weighted by the de Haar measure, which controls all the calculation. The method is well stablished; see, e.g.,  Refs.~\cite{bb96,nos1}.

\begin{figure}
\begin{center}
\includegraphics[width=7.0cm,height=3.2cm]{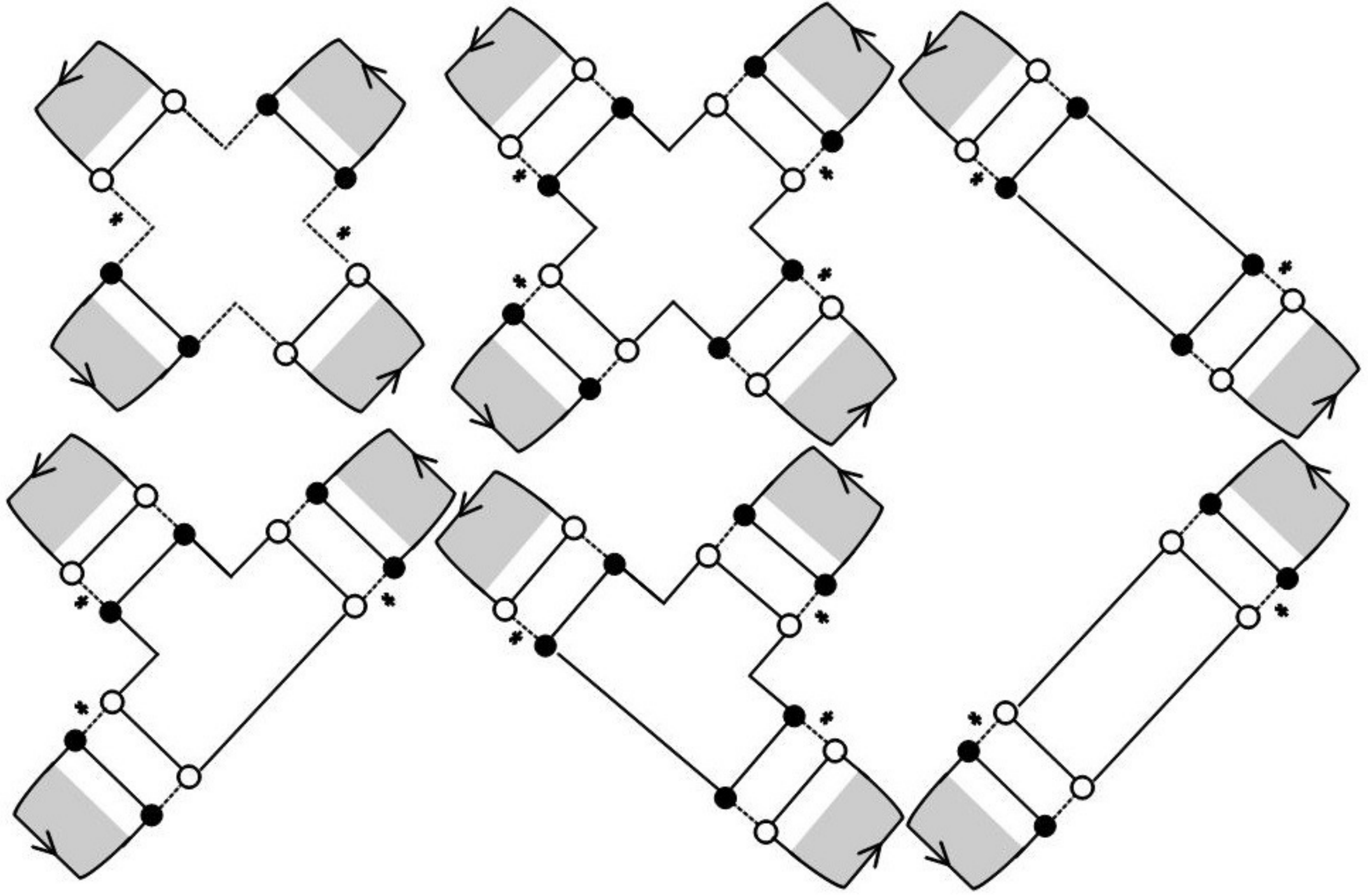}
\end{center}
\caption{Basic ladder diagrams which contribute to the main term of the average noise. The maximally crossed diagrams, which give the correction to the weak localization, are obtained through the process of crossing the arms of each one of the shown ladder diagrams.} \label{basic}
\end{figure}

To get to the result, we still need to calculate the average of ${\bf g_2}$. We follow \cite{nos1} to write
\begin{eqnarray}
\langle \mathbf{Tr} (tt^\dag)^2\rangle=\frac{N_{1}N_{2}\!
\left(N_1^2\!+\!N_1N_2\!+\!N_2^2\!-\!2N\!+\!1\!+\!\frac{4N\!-\!6}{\beta}\!+\!\frac{4}{\beta^2}\!\right)}
{\left( N-2+\frac{2}{\beta}\right)\left(
N-1+\frac{2}{\beta}\right)\left( N-1+\frac{4}{\beta}\right)},\nonumber
\label{segundocumulante2}
\end{eqnarray}
after using the same projectors given above.

It is instructive to obtain from (\ref{condutanciabeta}) and (\ref{segundocumulante2}) the first two terms in the perturbative
semiclassical expansion, which is valid provided $N_1,N_2\gg 1$, for the crossover in temperature of noise given by Eq.~\eqref{noiseprincipal}.
We obtain the following expression for average noise
\bes\be
\frac{\langle \mathcal{S}(k_B T,eV)\rangle}{4 k_B T G_0} =\langle \mathcal{S}_{SC}(k_B T,eV)\rangle+\langle \mathcal{S}_{WL}(k_B T,eV)\rangle,
\ee
with
\be
\langle \mathcal{S}_{SC}(k_B T,eV)\rangle= \frac{N_1N_2}{(N_1\!+\!N_2)^3} [{N_1^2\!+\!N_2^2}\!+\!{N_1 N_2}(1\!+\!F(\theta))],
\ee
as the main (semiclassical) part and
\ben
\langle \mathcal{S}_{WL}(k_B T,eV)\rangle\!&=&\!\frac{N_1N_2(1\!-\!2/\beta)}{(N_1+N_2)^4}\Bigl[{4N_1N_2}\nonumber
\\
&&+(N_1\!-\!N_2)^2(2\!-\!F(\theta))\Bigr],
\een\ees
as the first correction due to quantum interference.

If  $k_B T \gg eV$, the thermal or Johnson-Nyquist noise is dominant, and it has the form
\begin{eqnarray}
\frac{\langle \mathcal{S}(k_B T)\rangle}{4 k_B T G_0} =  \frac{N_1N_2}{N_1+N_2}\left[1+\left(1-\frac{2}{\beta}\right)\frac{1}{(N_1+N_2)}\right]. \label{noiseidealtemp}
\end{eqnarray}
This result gives $\langle \mathcal{S}(K_{B}T)\rangle=4 K_{B}T \langle G \rangle$. As expected, it is obtained by the dissipation-fluctuation theorem,
compatible with the limit $eV \rightarrow 0$ in Eq.~(\ref{noiseprincipal}). Thus, since the thermal noise only depends on the conductance, the result engenders no relevant information
about the temporal correlations between the electronic modes of the current in the left and right leads.

If $eV \gg k_B T$ we have the shot-noise power, which can be written as
\begin{eqnarray}
\frac{\langle \mathcal{S}(eV)\rangle}{2eV G_0}\! = \!\frac{N_1^2N_2^2}{(N_1\!+\!N_2)^3}\!\left[
{1}\!-\!\left(1\!-\!\frac{2}{\beta}\right)\frac{(N_1\!-\!N_2)^2}{N_1 N_2(N_1+N_2)} \right]. \label{noiseidealbias}
\end{eqnarray}
This result is in accordance with Refs.~\cite{beenakker,nos2}. The shot-noise power engenders important information on the temporal correlations between the electronic modes
in the left and right leads. Such informations are necessarily contained in the conductance. The shot-noise power plays a key role when the electronic transmission is chaotic.

It is of current interest to generalize the above result to non-ideal cavities, including effects of tunneling barriers.
Thus, let us focus on implementing a perturbative diagrammatic procedure to calculate the quantum noise for a chaotic cavity with
two barriers of arbitrary transparencies, with special attention to the leading quantum interference correction. In the presence of
barriers, the scattering matrix of the chaotic cavity is distributed according to the Poisson kernel \cite{mello,nos1}. The key idea here is to consider the diagrammatic technique used above, so the scheme is to map the ensemble average over the Poisson kernel to an effective problem with a random matrix belonging to one of the circular ensembles. If we follow \cite{bb96}, this is achieved by separating the average and the fluctuating part of the scattering matrix as follows, $S=\bar{S}+\delta S$, where $\bar S$ is a sub-unitary and diagonal matrix, $\bar{S}=\textrm{diag} (r_{1},r_{2})$ and $\delta S = T \left(1-RU\right)^{-1}UT$ is the fluctuating part. The matrix $U$ is orthogonal, unitary, or quaternionic, depending on the Dyson index being $\beta=1, 2$ or $4.$ It is also aleatory. The other matrices $T$ and $R$ are also diagonal, given by
$T={\rm diag}(i\sqrt{\Gamma_1} \,1_{N_{1}}, i\sqrt{\Gamma_2} \,1_{N_{2}})$ and $R={\rm diag}(\sqrt{1-\Gamma_1}\, 1_{N_{1}},\sqrt{1-\Gamma_2}\, 1_{N_{2}})$. Here
$r_1,r_2$ are the average of the reflection matrices in lead $1$ and $2$, respectively, and $\Gamma_1,\Gamma_2$ are the corresponding barrier values.

The polar decomposition of $\delta S$ allows to obtain the average as products of the fluctuating part of the scattering matrix. The average of the first and second cumulants can be written as  $\langle \mathbf{g}_{1} \rangle = \langle \mathbf{Tr} \left( C_{1} \delta S C_{2} \delta S^{\dagger} \right) \rangle$ and $\langle \mathbf{g}_{2} \rangle = \langle \mathbf{Tr} \left( C_{1} \delta S C_{2} \delta S^{\dagger} \right)^2\rangle$. The diagrammatic method emerges when we expand in power of $U$ the fluctuating part, to extract the average. The procedure generates two distinct classes of diagrams (diffusons and cooperons) having vertices which represent the indices of the unitary matrices. For further details, see, e.g., Ref.~\cite{nos1}.

The ladder diagrams for the general case in the presence of barriers are depicted in Fig.~\ref{basic}. The others, maximally crossed diagrams are obtained through the crossing of the arms of the ladder diagrams. Using the two classes of diagrams, Brouwer and Beenakker computed that
\begin{eqnarray}
{\langle\mathbf{Tr}(tt^\dag)\rangle}=\frac{G_1 G_2}{G_1+G_2}\!\left(\!1+\left(\!1-\frac{2}{\beta}\right)\frac{G_2 \Gamma_{1}+G_1 \Gamma_{2}}{(G_1+G_2)^2}\right), \label{tt10}
\end{eqnarray}
where $G_i=\Gamma_i N_i$, for $i=1,2.$

\begin{figure}
\begin{center}
\includegraphics[width=8.2cm,height=6.0cm]{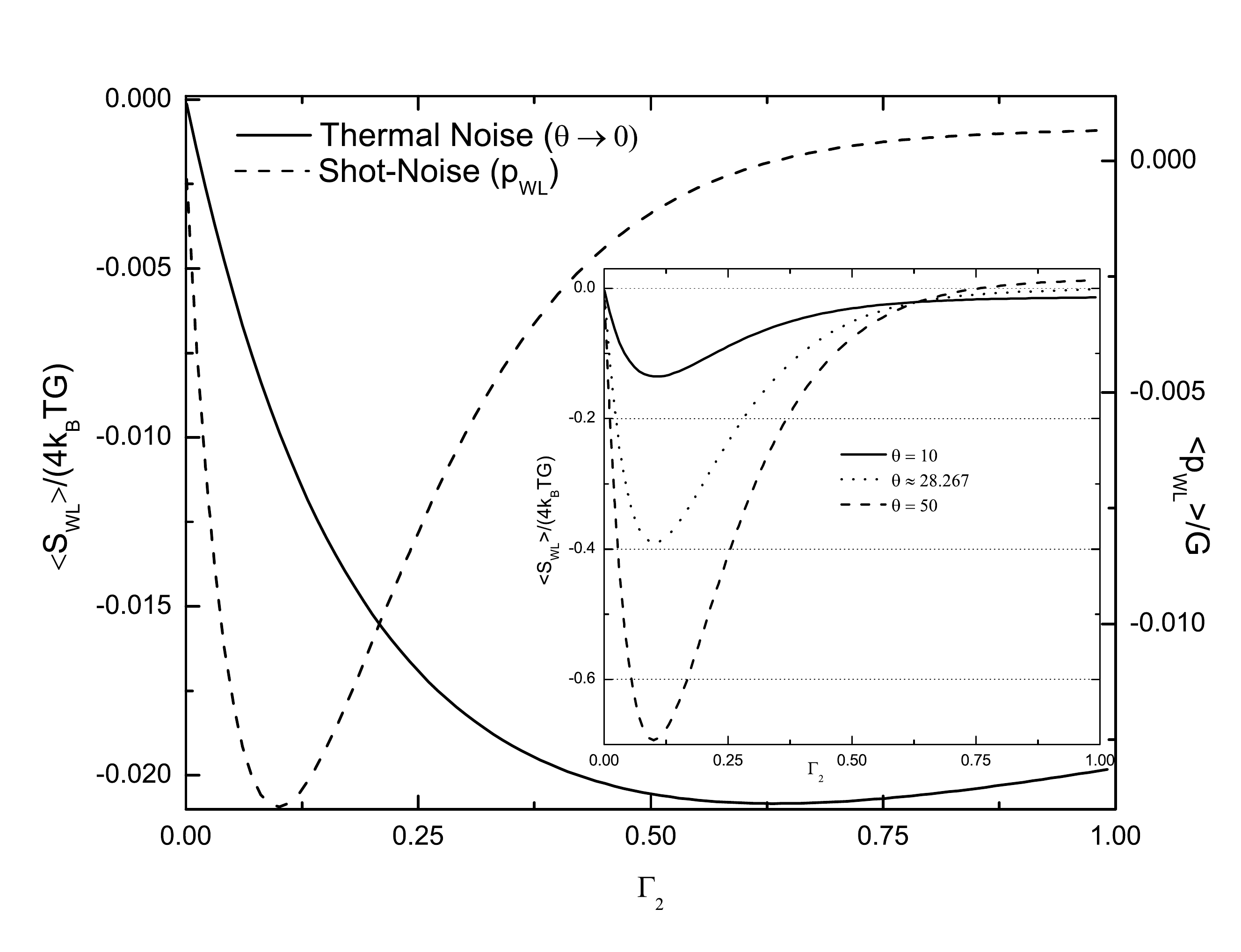}
\end{center}
\caption{Behavior of the main quantum correction to the noise as a function of $\Gamma_{2}$, for $\Gamma_{1}=0.95$. The solid and dashed curves of the main graphic show the weak localization contribution to the thermal and shot-noise power, respectively. In the inset one shows the noise behavior at three distinct temperatures, indicating the disappearance of the suppression-amplification effect as the temperature increases.} \label{swlt2}
\end{figure}

In Ref.~\cite{nos1} one depicts all the distinct topological diagrams which contribute to generate the average of the main quantum correction to the shot-noise. In the present study we use these diagrams for the case of equivalent channels to calculate the average in presence of barriers. We use the ladder diagrams (diffusons) of Fig.~\ref{basic}, 
to obtain the main term, after some algebraic manipulations,
\begin{eqnarray}
\langle\mathbf{Tr}(tt^\dag)^2\rangle_{\!SC}\!=\!\frac{G_1 G_2(G_1^3 \Gamma_{2}\!+\!2 G_1^2 G_2\!+\!2 G_1 G_2^2\!+\!G_2^3 \Gamma_{1})}{(G_1+G_2)^4}.
\end{eqnarray}
Next, we folow  \cite{bb96,nos1} to extract the maximally crossed diagrams (cooperons). They are obtained crossing the arms of the diagrams depicted in
Fig.~\ref{basic}. The procedure consists in getting all the topologically distinct diagrams, with their corresponding weights. After some algebra, we obtain the main quantum interference term in the form
\begin{eqnarray}
\langle\mathbf{Tr}(tt^\dag)^2\rangle_{WL}&\!\!\!=\!\!\!&2\!\left(\frac{2}{\beta}\!-\!1\!\right)\!\! \frac{G_1 G_2 (G_2 \Gamma_{1}\!+\!G_1 \Gamma_{2})\!\!}{(G_1+G_2)^6} \nonumber \\
&& \times \bigl[ 3G_1 G_2(G_1\!+\!G_2) -\!2G_1 G_2(G_1\Gamma_{2} \;\;\nonumber \\
&&+G_2\Gamma_{1})\! + \!G_1^3(2\Gamma_{2}\!-\!1)\!+\!G_2^3(2\Gamma_{1}\!-\!1)\bigr].\;\; \;\;\;\;\;\label{tt210}
\end{eqnarray}

We now use the former averages up to the weak localization correction in order to obtain the average of the crossover between thermal and shot-noise power under the presence of barriers.
The general expression has several terms, but it can be separated into two distinct contributions, one being the main term, semiclassical ($SC$), and the other the weak localization ($WL$) term. We make good use of the general structure of the result to write
\be
\mathcal{S}(k_B T,eV) \!=\! \mathcal{S}_{SC}(k_B T,eV)\! +\!\left(\frac{2}{\beta}-1\!\right)\mathcal{S}_{WL}(k_B T,eV),\nonumber
\ee
which allows investigating each one of the $SC$ and $WL$ terms separately.

The $S_{SC}$ term does not depend on the particular Dyson's index and it is given by Eq.~{\eqref{wt1}}. This result is equivalent to the one obtained in Ref.~\cite{nagaev},
\begin{widetext}
\be\label{wt1}
\frac{\langle \mathcal{S}_{SC}(k_B T,eV) \rangle}{4 k_B T G }=\frac{1}{(G_1+G_2)^2}\left\{{G_1G_2}(2+F(\theta))+
\frac{G_1^3\Gamma_{2}+G_2^3\Gamma_{1}}{G_1+G_2}(1-F(\theta))+\frac{G_1^3+G_2^3}{G_1+G_2}\;F(\theta)\right\},
\ee
\end{widetext}
where $G \equiv G_0 G_1 G_2/(G_1+G_2)$ is the main term of the conductance in the presence of barriers. We have studied the behavior of the $SC$ term as a function of the second barrier, for a given value of $\Gamma_{1}$, and for distinct values of the temperature. We have verified that there is no qualitative difference between the distinct curves.

The next term, $S_{WL}$, which adds quantum correction, depends on the Dyson's index and engenders surprising effects. It is the most important result of this work, and it is given by Eq.~{\eqref{wt2}}.
\begin{widetext}
\begin{eqnarray}\label{wt2}
\frac{\langle \mathcal{S}_{WL}(k_B T,eV) \rangle}{4k_B T G}&=& \left(G_1\Gamma_{2}+G_2\Gamma_{1}\right)
\biggl\{\frac{G_1G_2}{(G_1+G_2)^4}(2+F)-4\frac{G_1G_2}{(G_1+G_2)^5}\left(G_1\Gamma_{2}+G_2\Gamma_{1}\right)(1+F(\theta))\nonumber\\&&+
\frac{2G_1^3(2\Gamma_{2}-1)+2G_3^3(2\Gamma_{1}-1)}{(G_1+G_2)^5}(1+F(\theta))-\frac{G_1^3+G_2^3}{(G_1+G_2)^5}F(\theta)\biggr\}.
\end{eqnarray}
\end{widetext}
It represents corrections due to space and time correlations for the quantum dot in the presence of barriers, arbitrary number of channels, temperature and electromagnetic potential.

Interesting limits of the general result can be reported. For instance, if $eV \gg k_B T$ we have the shot noise power $\langle \mathcal{S}(eV)\rangle = 2 eV  G_0 \,\langle {\bf p} \rangle$,
as it was obtained in \cite{osipov} in the case of ideal contacts. On the other hand, if $k_B T \gg eV$ we have the thermal noise, also known as Johnson-Nyquist noise, obtained for the case of ideal contacts in Ref.~\cite{osipov}. In the presence of barriers we have
\be
\frac{\langle \mathcal{S}(k_B T)\rangle}{4 k_B T G} \!=\! 1\!+\!\left(1\!-\!\frac{2}{\beta}\right)\!\frac{\left(G_1\Gamma_{2}\!+\!G_2\Gamma_{1}\right)}{(G_1\!+\!G_2)^2}.\nonumber
\ee
This result suggests that the thermal noise is peculiar function for symmetric barriers: in this case we get $(G_1\Gamma_{2}\!+\!G_2\Gamma_{1})/(G_1\!+\!G_2)^2=1/(N_1+N_2)$. This means that up to the main quantum correction, there is no dependence of the barries, showing that the thermal noise only depends on the barriers through the main or semiclassical contribution which comes from the conductance, as the fluctuation-dissipation theorem would indicate, up to the first quantum correction to the thermal noise.

Another noteworthy effect which we observe is its linear suppression in the opaque limit. This limit was defined in
\cite{whitney} as $\Gamma_{i}\rightarrow 0$ and $N_{i} \rightarrow \infty$ with $G_i$ fixed. A similar effect for the weak-localization correction to
the conductance has a nice physical explanation in the semiclassical approach \cite{whitney}. In this limit we get to
\be
\frac{\langle\mathcal{S}(k_B T,eV) \rangle}{4 k_B T G}=\frac{{G_1G_2}(2+F(\theta))}{(G_1+G_2)^2}+\frac{G_1^3+G_2^3}{(G_1+G_2)^3}F(\theta) ,\nonumber
\ee
Our result shows that the weak localization contribution goes to zero linearly with $\Gamma_{i}$, even at finite temperature. In this sense, it generalizes the former result of Ref.~{\cite{whitney}}.

Let us now focus on the effects which appear from the crossover between the thermal and shot-noise power for $\beta=1$. In Fig. [\ref{swlt2}] we depict the behavior of the dominant quantum correction as a function of $\Gamma_{2}$, for $\Gamma_{1}$ fixed at the value $0.95$. The pure thermal noise is achieved at higher temperatures, much higher then the bias tension. It is depicted in Fig.~\ref{swlt2} with the solid curve, and we note that it is always negative. For the pure shot-noise power, one needs to set $T\to0$. It is also depicted in Fig.~\ref{swlt2} with the dashed curve. It shows that the suppression-amplification transition formerly reported in \cite{nos1,nos2} is still present, representing a sign change in the main quantum noise. Also, in the inset in Fig.~\ref{swlt2} one shows the possibility for the suppression-amplification effect to disappear as the temperature increases, leading to the main quantum noise with no sign change anymore. The critical temperature depends on the barriers and on the ratio between the number of open channels; for $\Gamma_1=0.95$ and $N_2/N_1=2/3$ one gets $\theta_c=28.267$. We remark that the critical temperature is achieved when the curve of the weak localization crosses the $\Gamma_2=1$ line at zero value, for $\Gamma_1$ fixed. The general procedure leads us to a transcendental equation of the form $\theta_c \coth(\theta_c)=\textrm{constant}$, which can always be solved to give the critical temperature.

\begin{figure}
\begin{center}
\includegraphics[width=8.2cm,height=8.0cm]{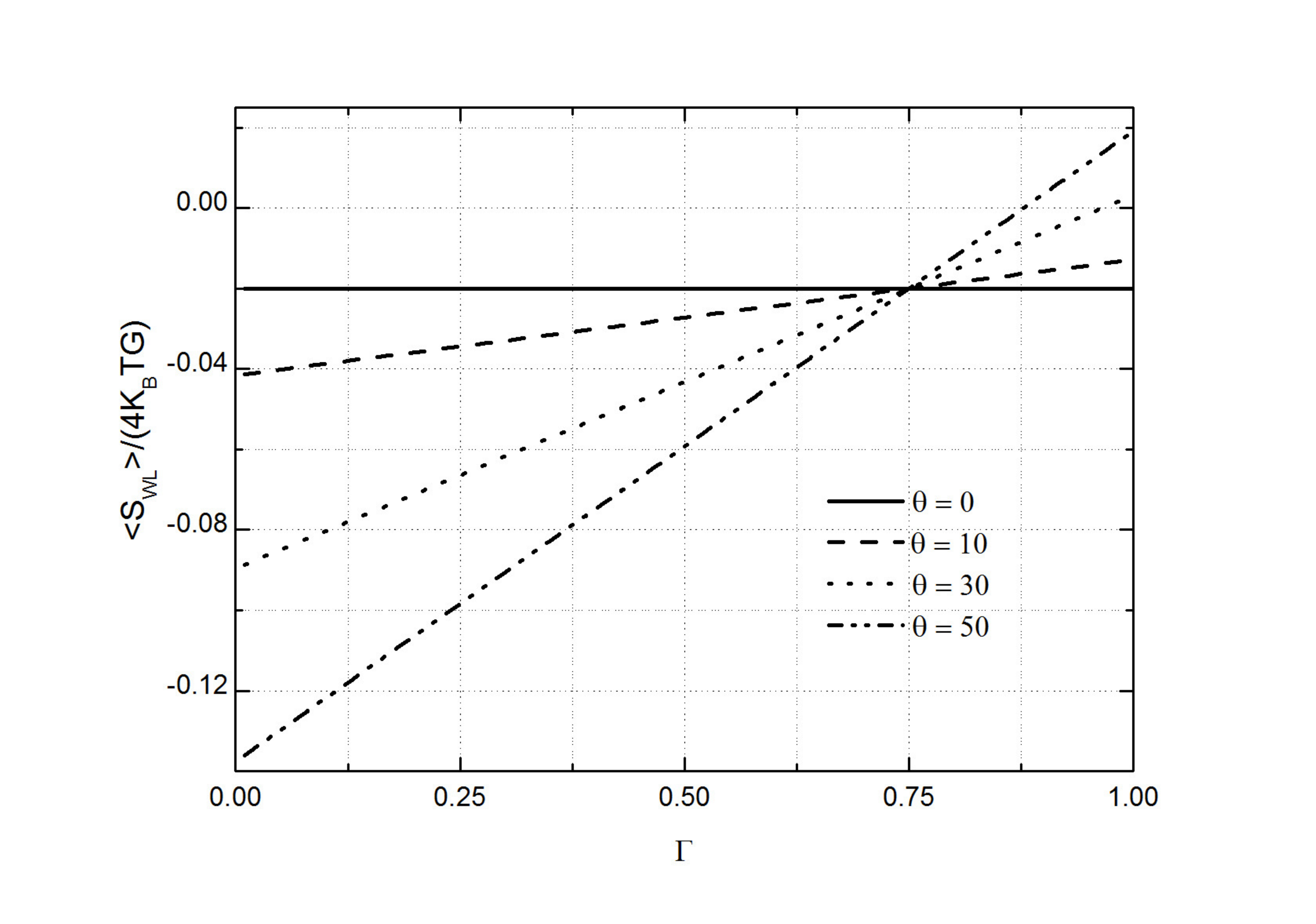}
\end{center}
\caption{Plot of the weak localization contribution to the noise, showing the appearance of a critical point at $\Gamma=0.75$, independently of the temperature.} \label{swlgamma}
\end{figure}

In the case of symmetric barriers, $\Gamma_{1}=\Gamma_{2}=\Gamma$, there appears the critical point at $\Gamma=3/4=0.75$. Independently of the number of open channels, the weak localization contribution to the shot-noise vanishes at this critical point, since $T\to0$ in this case. As it is shown in Fig.~\ref{swlgamma}, all the curves cross the critical point, independently of the temperature, including the case of $T\rightarrow\infty$. We note that the horizontal line in Fig.~\ref{swlgamma} is in accordance with the fluctuation-dissipation theorem,
since $\langle S_{WL}\rangle = 4 k_B T\langle G\rangle$. We also note that all the curves at finite temperature increase monotonically, and that the suppression-amplification effect reported in Refs.~\cite{nos1,nos2} is still visible, for $\theta>\theta_c$. In this sense, we are extending the suppression-amplification effect reported in \cite{nos1,nos2} to other scenarios.

\begin{figure}
\begin{center}
\includegraphics[width=8.2cm,height=7.0cm]{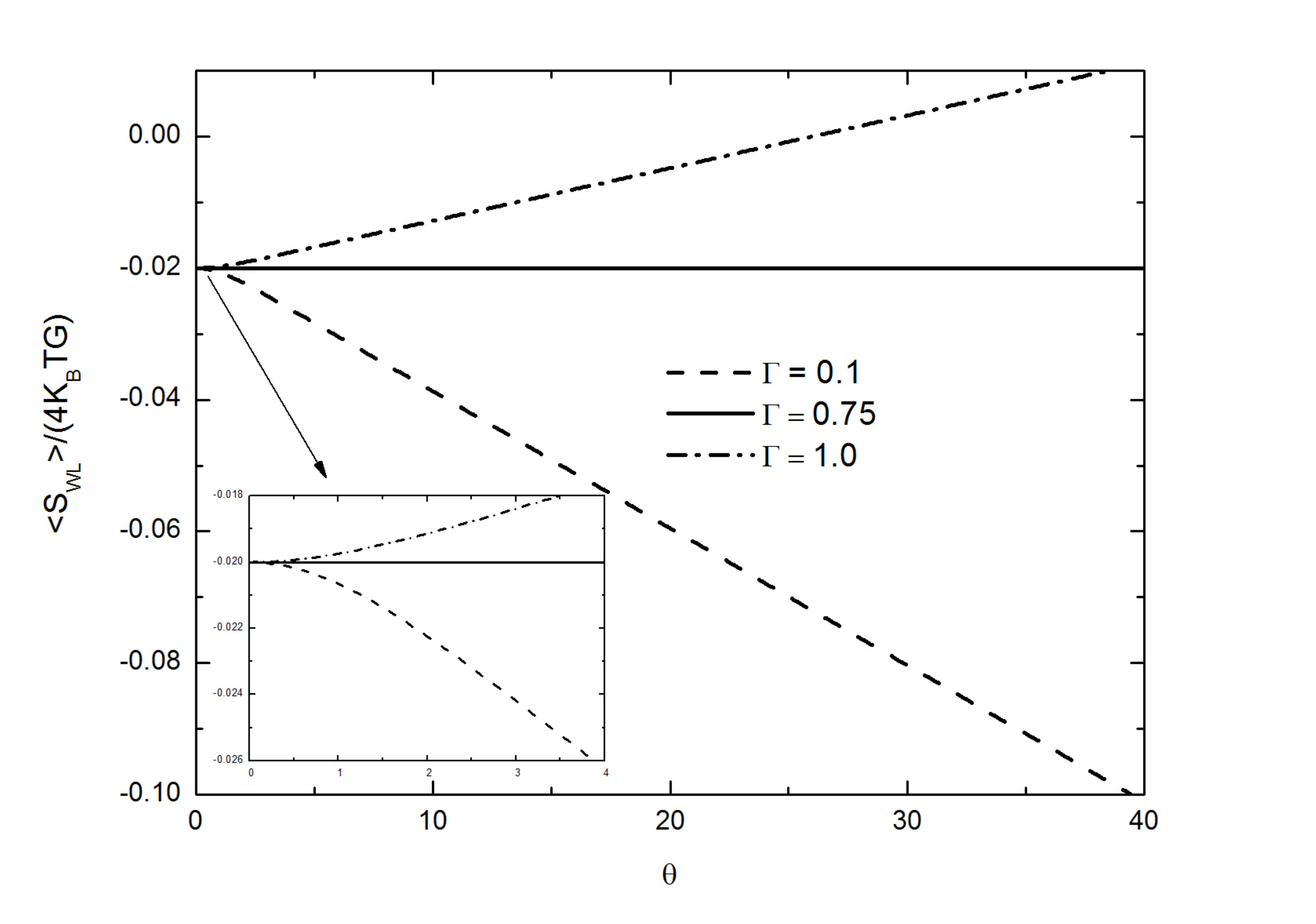}
\end{center}
\caption{Plot of the weak localization contribution to the noise, showing a change in the concavity of the curves as a function of the temperature, which occurs at the critical point, $\Gamma=0.75$.} \label{swltheta}
\end{figure}

Former result \cite{nos2} shows that there is a suppression-amplification effect which appears in the case of non-ideal contacts, at zero temperature. However, after implementing the diagrammatic calculation in the presence of barriers and temperature, the result changes significantly, showing that such suppression-amplification effect may disappear. Surprisingly, the anomalous change of the concavity of the main quantum correction to the noise occurs exactly at the critical point of the tunneling barriers, with $\Gamma=0.75$. In Fig.~\ref{swltheta} we plot the weak localization correction to the noise as a function of $\theta$, and there we see very clearly the change in its concavity, for different values of the tunneling barriers.
In this figure, we have fixed $\Gamma_{1}=\Gamma_{2}=\Gamma$, and we see that the function starts at the value $-1/(N_1+N_2)=-0.02$ (recall that we are using $N_1=30$ and $N_2=20$) at $\theta=0$, for any value of $\Gamma$, and it varies differently, for different values of $\Gamma$. The case $\Gamma=3/4$ is again special, generating a constant curve. It is noteworthy recalling result of Ref.~\cite{nos2}, which shows that the contribution due to the shot-noise power vanishes exactly at $\Gamma=3/4$, independently of the values of $N_1$ and $N_2$.

In summary, in this Letter we presented a quantitative analysis of the crossover of thermal to shot-noise power in a chaotic cavity with non-ideal contacts. 
We studied competition between tunneling and interference among the electronic carriers. Introducing the thermal effects, which contributes to modify the Fermi-Dirac distribution in the two reservoirs, we described the crossover between the shot noise and the thermal noise. Surprisingly, we have shown that the tunneling barrier effects are nontrivial, at least from the phenomenological point of view, leading to critical curves and points which can be of direct interest to future experiments, such as the one of Ref.~\cite{gustavson}. The calculations were implemented with the diagrammatic method, used to perform the integration in the unitary group for the case of equivalent channels. The main results of the paper are valid for each one of the Wigner-Dyson's ensembles. In particular, we have shown that the suppression-amplification effect for the noise described in \cite{nos2} do contribute even at finite temperature, but it disappears at a certain temperature. Moreover, the main quantum correction to the noise has an anomalous thermal behavior when one let the barriers vary.

The results obtained in this work explore several cases of practical importance, including the opaque limit. They can be tested experimentally and, in this sense, it would be interesting to prepare experiment to search in particular for the critical point related to the behavior of the weak localization with the tunneling barriers described in the present study. The present study can be used to extend the result of Ref.~\cite{pekola} to the case where the crossover of thermal and shot-noise power is obtained with the same methodology above, but now at finite frequency. We will report on this elsewhere.

We would like to thank CAPES and CNPq for financial support. ALRB and JGGSR also thank Professor A. M. S. Mac\^edo for introducing them to the subject explored in this work.


\end{document}